\documentclass[prd,preprint,nofootinbib,12pt]{revtex4}
\usepackage{amsfonts}
\usepackage{amsmath}
\usepackage{amssymb}
\usepackage{graphicx}
\usepackage{hyperref}
\usepackage[T1]{fontenc}
\usepackage{xcolor}
\usepackage{amsthm}
\usepackage{enumitem}

\newcommand{\dd}{{\rm{d}}} 

\newcommand{\im}{\mathrm{i}}
\newcommand{\Ree}{\mathrm{Re\,}}
\newcommand{\Imm}{\mathrm{Im\,}}

\newtheorem{theorem}{Theorem}

\begin{document}

\title{Uniqueness of stationary axisymmetric type D black holes with non-aligned electromagnetic field}

\author{Hryhorii Ovcharenko}
\email{hryhorii.ovcharenko@matfyz.cuni.cz}
\affiliation{Charles University, Faculty of Mathematics and Physics,
Institute of Theoretical Physics,
V~Hole\v{s}ovi\v{c}k\'ach 2, 18000 Prague 8, Czechia}

\author{Ji\v{r}\'i Podolsk\'{y}}
\email{jiri.podolsky@matfyz.cuni.cz}
\affiliation{Charles University, Faculty of Mathematics and Physics,
Institute of Theoretical Physics,
V~Hole\v{s}ovi\v{c}k\'ach 2, 18000 Prague 8, Czechia}

\date{\today}
\begin{abstract}
    We demonstrate the uniqueness of the spacetimes recently found by us in [H.~Ovcharenko and J.~Podolsk\'{y}, Phys.~Rev.~D {\bf 112} (2025) 064076]. First, we prove that the conformal-to-Carter metric ansatz we used therein is the only possible for stationary axisymmetric geometries that are of Weyl type~D, with geodesic and shear-free principal null directions (PNDs) which are orthogonal to polar directions, and whose specific 1-form $\theta$ is closed. Because this result is general, without employing any field equations, such conformal-to-Carter metric may find interesting applications also in various alternative theories of gravity. Then, we show that in the Einstein-Maxwell theory the only non-trivial electrovacuum solution for the conformal-to-Carter metric with the fully non-aligned and non-null electromagnetic field is the Ovcharenko-Podolsk\'{y} class found in 2025. Complementarily, the only solution with the double-aligned and non-null electromagnetic field is the Pleba\'nski-Demia\'nski class found in 1976.
\end{abstract}
\maketitle

\tableofcontents

\newpage

\section{Introduction}

Investigation of exact type D spacetimes has always been intriguing. Such vacuum spacetimes, and those with an electromagnetic field aligned with both the double-degenerate principal null directions (PNDs) of the Weyl tensor, were thoroughly investigated \cite{Carter1968,Kinnersley1969,Plebanski:1976gy,Debever1983,Griffiths2006,Podolsky2021,Ovcharenko2025_1} (see \cite{Stephanietal:2003,Griffiths2009} for a review). On the other hand, the case of a \emph{non-aligned} electromagnetic field has not been studied widely because in this case the field equations become extremely complicated.

A breakthrough in this topic was recently achieved in the work by Van den Bergh and Carminati \cite{VandenBergh2020}, in which the authors were able to find the most general spacetime of algebraic type D with a non-aligned electromagnetic field, but only in the static (non-twisting) case. Moreover, this work is rather mathematical and does not concentrate on physical interpretation. Motivated by this, we generalized the Van den Bergh-Carminati solution to incorporate the twist \cite{Ovcharenko2025}, and elaborated on the physical interpretation of the novel large class of spacetimes. As shown in \cite{Ovcharenko2025,Podolsky:2025tle}, the new key ingredient distinguishing this class from the Pleba\'{n}ski-Demia\'{n}ski one \cite{Plebanski:1976gy}, namely that the electromagnetic field is non-aligned, can be interpreted as an external uniform (electro-)magnetic field of the Bertotti-Robinson type, into which the (possibly rotating and charged) black hole is immersed.

Even though the new class has already found many physical applications \cite{Zeng:2025olq,Wang:2025vsx,Zeng:2025tji,Wang:2025bjf,Astorino:2025lih,Ali:2025beh,Vachher:2025jsq,Zhang:2025ole,Gray2025}, the solution presented in \cite{Ovcharenko2025} was not proven to be the most general family of type D solutions of the Einstein-Maxwell equations with a non-aligned electromagnetic field. It implicitly contains several crucial assumptions, namely:

(i) The spacetime is stationary and axisymmetric, implying (in adapted frame)~${\Phi_0=\Phi_2}$.

(ii) The metric ansatz is given by
\begin{align}
    \dd s^2=\dfrac{1}{\Omega^2}\Big[-\dfrac{Q}{\rho^2}(\dd \eta-p^2\dd \sigma)^2
    +\dfrac{P}{\rho^2}(\dd \eta+q^2\dd \sigma)^2
    +\dfrac{\rho^2}{Q}\dd q^2+\dfrac{\rho^2}{P}\dd p^2\Big],\label{gen_metr}
\end{align}

where
\begin{align}
    \rho^2=q^2+p^2,\qquad
    Q(q)\,, \quad P(p)\,, \quad\Omega(q,p)\,.\label{rho-Q-P-Omega_assump}
\end{align}

(iii) The non-aligned part of the electromagnetic field has the special form
\begin{align}
    \Phi_0=\dfrac{c}{\Omega} \dfrac{\sqrt{P Q}}{q+\im \,p},\label{phi_0_assump}
\end{align}

where $c=\mathrm{const.}$ (complex).

These assumptions are quite strong, but were necessary for the explicit integration of the field equations. Our present work aims to relax these assumptions, in an attempt to generalize this class. For this, we have to ask two questions:
\vspace{2mm}

1) What are the most general \emph{algebraic} properties that imply that the metric takes the form (\ref{gen_metr})?

2) Is the solution to the Einstein-Maxwell equations with (\ref{rho-Q-P-Omega_assump}), (\ref{phi_0_assump}) the \emph{most general} non-aligned solution, or not?
\vspace{2mm}

We answer these questions by proving the following two theorems, namely:

\begin{theorem}\label{th_1}
    Consider a 4-dimensional spacetime that satisfies the conditions:
    \begin{enumerate}[label=\textnormal{(\roman*)}]
        \item \label{th1} It is stationary and axisymmetric (with non-null Killing vectors).
        \item \label{th2} It is of algebraic type D.
        \item \label{th3} Both PNDs $\mathbf{k}$ and $\mathbf{l}$ of the Weyl tensor are geodesic and shear-free.
        \item \label{th4} Both PNDs $\mathbf{k}$ and $\mathbf{l}$ of the Weyl tensor are orthogonal to polar direction.\footnote{This is specified below Eq.~\eqref{gen_ax_sym_0}.}
        \item \label{th5} The 1-form $\theta=\mu\,\mathbf{k}-\rho\,\mathbf{l}-\pi\,\mathbf{m}+\tau\,\bar{\mathbf{m}},$ with $\mathbf{k}$ and $\mathbf{l}$ being PNDs, is closed.
    \end{enumerate}
    Then it is given by the conformal-to-Carter metric (\ref{gen_metr}), (\ref{rho-Q-P-Omega_assump}).
\end{theorem}

This theorem will be proven in Section \ref{sec_gen_metr}.

\begin{theorem}\label{th_2}
    {The most general solutions to the Einstein-Maxwell equations for the conformal-to-Carter metric (\ref{gen_metr}), (\ref{rho-Q-P-Omega_assump}) are:}
    \begin{enumerate}[label=\textnormal{(\roman*)}]
        \item  With the double-aligned and non-null electromagnetic field it is the Pleba\'{n}ski-Demia\'{n}ski spacetime \cite{Plebanski:1976gy}.
        \item  With the fully non-aligned and non-null electromagnetic field it is the Ovcharenko-Podolsk\'{y} spacetime \cite{Ovcharenko2025}.
    \end{enumerate}
\end{theorem}
This theorem will be proven in Section \ref{sec_uniq}.

\newpage

Before we proceed with the proof of these two theorems, we would like to comment on the assumptions of Theorem \ref{th_1}.

Assumption (i) is quite natural for black holes, as many known realistic black hole spacetimes are stationary axisymmetric, and the Killing vectors are not null everywhere, except on the horizon. Moreover, the case of null Killing vectors can be obtained as the near-horizon limit of the resulting spacetimes, as we will discuss during the proof.

Assumption (ii), requiring the spacetime to be of algebraic type D, is also quite natural because main black hole spacetimes, such as Schwarzschild or Kerr, are of this type.

Assumption (iii) has to be commented in more detail. For vacuum spacetimes, the assumptions (ii) and (iii) are \emph{equivalent} because of the Goldberg-Sachs theorem. Also, they are equivalent in electrovauum spacetimes with an aligned electromagnetic field due to the Kundt-Tr\"{u}mper theorem \cite{Kundt1963}. However, generally, the condition (iii) does not follow from (ii), and has to be assumed as a distinct additional condition.

Assumptions (iv) and (v) may seem obscure and strange, as they do not have a direct algebraic/geometric meaning. However, as we will show during the proof, they are required to obtain the unique metric (\ref{gen_metr}), (\ref{rho-Q-P-Omega_assump}).

Also, we wish to comment on the frame freedom, and the related uniqueness. This question is motivated by the fact that in Theorem \ref{th_1} we assume the conditions (iii) and (v) that involve optical scalars. In particular, they require us to choose a specific null frame such that the form $\theta$ defined in (v) is closed. But it can be shown that if $\mathbf{k}$ and $\mathbf{l}$ are the PNDs of the type D Weyl tensor, then $\theta$ is invariant with respect to null rotations and boosts.

Indeed, for a given spacetime, one has the freedom to conduct two null rotations, one spin rotation, and one boost. We will use the two null rotations to choose such $\mathbf{k}$ and $\mathbf{l}$ that they coincide with the PNDs of the type D Weyl tensor. Then we are left with the boost and the spin rotation that transform the null frame as
\begin{align}
    \mathbf{k}'=B\,\mathbf{k},~~~\mathbf{l}'=B^{-1}\mathbf{l},~~~\mathbf{m}'=e^{\im \Theta}\mathbf{m},~~~\bar{\mathbf{m}}'=e^{-\im \Theta}\bar{\mathbf{m}}\label{eq_3}.
\end{align}

During this transformation, the optical scalars transform in such a way that
\begin{align}
    &\kappa'=B^2e^{\im \Theta}\kappa,~~~\sigma'=Be^{2\im \Theta}\sigma,~~~\lambda'=B^{-1}e^{-2\im \Theta}\lambda,~~~\nu'=B^{-2}e^{-\im \Theta}\nu,\label{eq_4}\\
    &\rho'=B\rho,~~\mu'=B^{-1}\mu,~~\tau'=e^{\im \Theta}\tau,~~\pi'=e^{-\im \Theta}\pi.\label{eq_5}
\end{align}

We notice that if the PNDs are geodesic and shear-free, $\kappa=\nu=0=\sigma=\lambda$, then after the boost and spin rotations (\ref{eq_4})--(\ref{eq_5}), they remain geodesic and shear-free, $\kappa'=\nu'=0=\sigma'=\lambda'$. Moreover, the 1-form $\theta$ defined in (v) is also invariant under these transformations, $\theta'=\theta$. This means that during our proof, without the loss of generality we can choose \emph{any} null frame with $\mathbf{k}$ and $\mathbf{l}$ both co-directed with the PNDs of the Weyl tensor, because the assumptions (iii) and (v) (that are the only ones that involve the optical scalars) of Theorem \ref{th_1} are independent of the transformation \eqref{eq_3}.

\section{Proof of Theorem 1: Generality of the metric ansatz}\label{sec_gen_metr}

Let us consider the most general stationary axisymmetric metric. It can be written as\footnote{Actually, $N^2$ could be also negative, which is equivalent to considering $Q<0$ in the metric (\ref{gen_ax_sym}).}
\begin{align}
     \dd s^2=-N^2 \dd t^2+g_{\varphi\varphi} (\dd \varphi-\omega \dd t)^2+\dfrac{\dd r^2}{A}+g_{xx}\, \dd x^2,\label{gen_ax_sym_0}
\end{align}
where all the metric functions depend only on $r$ and $x$.\footnote{Usually, for stationary axisymmetric spacetimes, instead of the coordinate $x$ one uses the polar coordinate $\vartheta$ defined via the relation ${x=\cos\vartheta}$. However, for the simplicity of calculations, here we use $x$.} This metric thus clearly contains two Killing vectors, namely $\partial_{t}$ and $\partial_{\varphi}$. In addition, there are two spatial directions, namely the \emph{``radial'' direction}~$\partial_r$, and the \emph{``polar'' direction}~$\partial_{x}$.

One could employ these metric functions, but it is better to introduce new ones, namely\footnote{Here we introduce a function $\rho^2$ that can be chosen arbitrarily. We will use this freedom later.}
\begin{align}
    Q&=A\rho^2\sqrt{g_{xx}}\,\dfrac{\sqrt{N^2-\omega^2\,g_{\varphi\varphi}}}{\sqrt{A \,g_{xx}-1}},\nonumber\\
    P&=\dfrac{\rho^2}{\sqrt{g_{xx}}}\,\dfrac{\sqrt{N^2-\omega^2\,g_{\varphi\varphi}}}{\sqrt{A \,g_{xx}-1}},\nonumber\\
    v&=\dfrac{\sqrt{A\,N^2 \,g_{xx}\,g_{\varphi\varphi}}-A \,\omega \,g_{xx}\,g_{\varphi\varphi}}{A\,g_{xx}(N^2-\omega^2\,g_{\varphi\varphi})},\label{new_funcs}\\
    u&=\dfrac{g_{\varphi\varphi}\,\omega-\sqrt{A \,N^2\,g_{xx}\,g_{\varphi\varphi}}}{N^2-\omega^2\,g_{\varphi\varphi}},\nonumber\\
    \Omega^2&=\dfrac{1}{\sqrt{g_{xx}}}\,\dfrac{\sqrt{A\, g_{xx}-1}}{\sqrt{N^2-\omega^2\,g_{\varphi\varphi}}}.\nonumber
\end{align}
Let us also relabel the coordinates as
\begin{align}
    t\equiv \eta,~~~r\equiv q,~~~x\equiv p,~~~\varphi\equiv\sigma.\label{relab}
\end{align}
Combining (\ref{gen_ax_sym_0}) with (\ref{new_funcs})--(\ref{relab}), the metric takes the form
\begin{align}
    \dd s^2=\dfrac{1}{\Omega^2}\Big[-\dfrac{Q}{\rho^2}(\dd \eta-v\,\dd \sigma)^2+\dfrac{P}{\rho^2}(\dd \eta+u\, \dd \sigma)^2+\dfrac{\rho^2}{Q}\dd q^2+\dfrac{\rho^2}{P}\dd p^2\Big]. \label{gen_ax_sym}
\end{align}

Notice that this metric is \emph{much more general than} the one presented in \eqref{gen_metr}, because here $Q$ and $P$ are \emph{general} functions of \emph{both} the coordinates $q$ and $p$, while in \eqref{gen_metr} $Q$ is a function of $q$, and $P$ is a function of $p$ only. Also, the functions $u$ and $v$ are not simply $p^2$ and~$q^2$, respectively, but they are general functions $u=u(q,p)$ and $v=v(q,p)$. In addition, in the above construction, we introduced a new \emph{arbitrary} function $\rho^2$. To simplify further calculations, let us express it as
\begin{align}
    \rho^2(q,p)\equiv u(q,p)+v(q,p).\label{eq_10}
\end{align}

Now, let us proceed to the proof of Theorem \ref{th_1}. According to it, we have to prove that the requirement of having a type~D spacetime with the geodesic and shear-free PNDs implies $Q=Q(q)$, $P=P(p)$, and also ${u=q^2}$, ${v=p^2}$.

    To do this, let us choose some orthonormal tetrad, for example
\begin{align}
    \mathbf{e}_0&=\dfrac{1}{\sqrt{Q}}\dfrac{\Omega}{\rho}(u\,\partial_{\eta}-\partial_{\sigma}),~~~\mathbf{e}_1=\dfrac{\Omega}{\rho}\sqrt{Q}\,\partial_q,\nonumber\\
    \mathbf{e}_3&=\dfrac{1}{\sqrt{P}}\dfrac{\Omega}{\rho}(v\,\partial_{\eta}+\partial_{\sigma}),~~~\mathbf{e}_2=\dfrac{\Omega}{\rho}\sqrt{P}\,\partial_p.\label{tetrad}
\end{align}
Using this, we define the null tetrad
\begin{align}
    \mathbf{k}&=\dfrac{1}{\sqrt{2}}(\mathbf{e}_0+\mathbf{e}_1),~~~\mathbf{l}=\dfrac{1}{\sqrt{2}}(\mathbf{e}_0-\mathbf{e}_1),\nonumber\\
    \mathbf{m}&=\dfrac{1}{\sqrt{2}}(\mathbf{e}_3+\im \mathbf{e}_2),~~~\bar{\mathbf{m}}=\dfrac{1}{\sqrt{2}}(\mathbf{e}_3-\im \mathbf{e}_2).\label{null_tetr_1}
\end{align}
This null tetrad is quite special because for it the relations\footnote{Generally, such $\mathbf{k}$ and $\mathbf{l}$ do not correspond to the PNDs of the Weyl tensor.}
\begin{align}
\alpha=\beta,~~~\pi=\tau,~~~\rho=\mu,~~~\epsilon=\gamma,~~~\kappa=\nu,~~~\sigma=\lambda
\end{align}
are automatically satisfied. This allows us to calculate the Weyl scalars
\begin{align}
    \Psi_0&=C_{\alpha\beta\gamma\delta}\,k^{\alpha}m^{\beta}k^{\gamma}m^{\delta},\nonumber\\
    \Psi_1&=C_{\alpha\beta\gamma\delta}\,k^{\alpha}l^{\beta}k^{\gamma}m^{\delta},\nonumber\\
    \Psi_2&=C_{\alpha\beta\gamma\delta}\,k^{\alpha}m^{\beta}\bar{m}^{\gamma}l^{\delta},\label{Weyl_scalars}\\
    \Psi_3&=C_{\alpha\beta\gamma\delta}\,l^{\alpha}k^{\beta}l^{\gamma}\bar{m}^{\delta},\nonumber\\
    \Psi_4&=C_{\alpha\beta\gamma\delta}\,l^{\alpha}\bar{m}^{\beta}l^{\gamma}\bar{m}^{\delta}.\nonumber
\end{align}

Direct calculation for the metric (\ref{gen_ax_sym}) gives us the following relations
\begin{align}
    \Psi_4=\bar{\Psi}_0,~~~~~~\Psi_3=\Psi_1.\label{psis_relation}
\end{align}
These specific relations are the consequence of the assumed symmetries, see \cite{Ovcharenko2025naked,Tanatarov2014}. Notice also that in these works it was shown that the algebraically special stationary axisymmetric spacetimes may only be of algebraic types II, D, N, or O.

Now, let us move to the task of finding the PNDs. The general procedure (see, e.g., Sec.~2 in \cite{Griffiths2009} for more details) is to perform a null rotation, keeping the null vector $\mathbf{l}$ fixed,
\begin{align}
    \mathbf{k}'=\mathbf{k}+K \bar{\mathbf{m}}+\bar{K}\mathbf{m}+K\bar{K}\,\mathbf{l},~~~\mathbf{l}'=\mathbf{l},~~~\mathbf{m}'=\mathbf{m}+K\,\mathbf{l}\,.\label{null_rot}
\end{align}
By this rotation, the Weyl scalar $\Psi_0$ transforms as
\begin{align}
    \Psi_0'=\Psi_0+4K\Psi_1+6K^2
    \Psi_2+4K^3\Psi_3+K^4 \Psi_4.\label{PND_eq}
\end{align}

The PND is such a direction $\mathbf{k}'$ that $\Psi_0'=0$. Generally, this equation has up to 4 distinct complex solutions $K$, but for type D black holes, there are 2 double-degenerate solutions.

Now, let us inspect what restrictions arise from the condition that the spacetime is of algebraic type D. Using the relations (\ref{psis_relation}) we obtain the equation for the PNDs in the form
\begin{align}
    (\Psi_0+\bar{\Psi}_0K^4)+4K(1+K^2)\Psi_1+6K^2
    \Psi_2=0\,.
\end{align}
This equation has two degenerate roots, that is ${\bar{\Psi}_0(K-K_+)^2(K-K_-)^2=0}$, \emph{only if}
\begin{align}
    \Psi_0=\bar{\Psi}_0,\qquad\Psi_0(3\Psi_2-\Psi_0)=2\Psi_1^2\,.\label{typ_D_conds}
\end{align}
If these conditions hold, the corresponding two solution for $K$ are given by
\begin{align}
    K_{\pm}=\pm \sqrt{\dfrac{\Psi_1^2}{\Psi_0^2}-1}-\dfrac{\Psi_1}{\Psi_0},
\end{align}
which also satify an important relation
\begin{align}
    K_{+}K_-=1.\label{K_pm_rel}
\end{align}

Now let us substitute \eqref{null_tetr_1} into \eqref{null_rot}, which yields
\begin{align}
    \mathbf{k}'=\dfrac{1}{\sqrt{2}}\Big[\mathbf{e}_0\,(1+K\bar{K})+\mathbf{e}_1\,(1-K\bar{K})+\mathbf{e}_3\,(K+\bar{K})+\im\, \mathbf{e}_2\,(K-\bar{K})\Big].
\end{align}

The condition (iv) in Theorem \ref{th_1} requires PNDs to be orthogonal to the \emph{polar direction} ${\partial_x\equiv\partial_p}$, which in view of (12) reads ${\mathbf{k}'\cdot \mathbf{e}_2=0}$. This means that $K$ has to be a \emph{real} quantity, ${K=\bar{K}}$, and $\mathbf{k}'_+$ becomes
\begin{align}
     \mathbf{k}'_+=\dfrac{1}{\sqrt{2}}\Big[\mathbf{e}_0\,(1+K_+^2)+\mathbf{e}_1\,(1-K_+^2)+2\mathbf{e}_3\,K_+\Big].
\end{align}
However, we can also perform a boost \eqref{eq_3} in the ${\mathbf{k}'_+ ,\mathbf{l}}$ directions with a specific boost parameter ${B_+=\dfrac{1}{1-K_+^2}}$, obtaining
\begin{align}
    \mathbf{k}_+''=\dfrac{1}{\sqrt{2}}\Big[\mathbf{e}_0\,\cosh\alpha+\mathbf{e}_3\,\sinh\alpha+\mathbf{e}_1\Big],
\end{align}
where
\begin{align}
    \tanh\alpha\equiv\dfrac{2K_+}{1+K_+^2}.
\end{align}

Analogously $\mathbf{k}'_-$ is given by
\begin{align}
     \mathbf{k}'_-=\dfrac{1}{\sqrt{2}}\Big[\mathbf{e}_0\,(1+K_-^2)+\mathbf{e}_1\,(1-K_-^2)+2\mathbf{e}_3\,K_-\Big],
\end{align}
but using the relation (\ref{K_pm_rel}), one obtains
\begin{align}
    \mathbf{k}'_-=\dfrac{1}{\sqrt{2}\,K_+^2}\Big[\mathbf{e}_0\,(1+K_+^2)+\mathbf{e}_1\,(K_+^2-1)+2\mathbf{e}_3\,K_+\Big].
\end{align}
Finally, by performing a boost (\ref{eq_3}) in the ${\mathbf{k}'_- ,\mathbf{l}}$ directions with a specific boost parameter $B_-=\dfrac{K_+^2}{1-K_+^2}$, one obtains
\begin{align}
    \mathbf{k}''_-=\dfrac{1}{\sqrt{2}}\Big[\mathbf{e}_0\,\cosh\alpha+\mathbf{e}_3\,\sinh\alpha-\mathbf{e}_1\Big].
\end{align}

After such rescalings, we have two null vectors $\mathbf{k}''_{\pm}$ that both are PNDs of the Weyl tensor, and these null vectors are properly normalized as $\mathbf{k}''_{+}\cdot\mathbf{k}''_-=-1$. We can thus associate a \emph{new null frame} with these vectors $\tilde{\mathbf{k}}=\mathbf{k}''_+, \tilde{\mathbf{l}}=\mathbf{k}''_-$. As the additional (complex) spatial vectors, we can use $\mathbf{m}$ and $\bar{\mathbf{m}}$ already introduced in (\ref{null_tetr_1}).

This new tetrad frame $(\tilde{\mathbf{k}},~\tilde{\mathbf{l}},~\mathbf{m},~\bar{\mathbf{m}})$ possesses interesting properties. Namely, the optical scalars are related as
\begin{align}
    \alpha&=\beta,~~~\tau=\pi,~~~\rho=\mu,~~~\epsilon=\gamma,~~~\kappa=\nu,~~~\sigma=\lambda,\\
    \tau&+\bar{\tau}=2(\alpha+\bar{\alpha}),~~~\rho-\bar{\rho}=2(\epsilon-\bar{\epsilon}).
\end{align}
Moreover, because the spacetime is stationary and axisymmetric, the action of the differential operators $D+\Delta$ and $\delta+\bar{\delta}$ on any optical scalar is zero.

Thus, all the conditions that have to be satisfied for the Theorem \ref{th_1} are
\begin{align}
    &\kappa=0=\nu,~~~\sigma=0=\lambda,\label{albeg_conds_1}\\
    &\alpha=\beta,~~~~\tau=\pi,~~~~\rho=\mu,~~~~\epsilon=\gamma,\\
    &\tau+\bar{\tau}=2(\alpha+\bar{\alpha}),~~~~\rho-\bar{\rho}=2(\epsilon-\bar{\epsilon}),\\
    &\dd\,\theta=0~~~\mathrm{for}~~~\theta=\mu\, \mathbf{k}-\rho\, \mathbf{l}-\pi\, \mathbf{m}+\tau\, \bar{{\mathbf{m}}}.\label{albeg_conds_4}
\end{align}
and that the action of the operators ${D+\Delta}$ and ${\delta+\bar{\delta}}$ on all optical scalar is zero.
From these conditions, one can find the constraints on the metric functions and to find the boost parameters by a direct computation. However, this is not necessary because it was already done by Debever, Kamran, and McLenaghan in their work \cite{Debever1984}.

Namely, in this work the authors were interested in finding the most general electrovacuum spacetimes of algebraic type D with the aligned electromagnetic field. By analyzing the Newman-Penrose field equations, they found that they \emph{imply} (\ref{albeg_conds_1})--(\ref{albeg_conds_4}). These are Eqs.~(3.1), (3.3)--(3.5) and (3.14) in \cite{Debever1984}. Here we consider only the case of timelike group orbits, so their parameter $f$ is taken to be $1$.

These strong conditions on the optical scalars then allowed them to use the Frobenius theorem and find the metric satisfying these properties. It is given by equation (2.5a) in \cite{Debever1984}. This metric can be rewritten in the form (see section 2.2.1 in \cite{Ovcharenko2024thesis} for more details):
\begin{align}
    \dd s^2=\dfrac{1}{\Omega^2}\Big[-\dfrac{Q}{\rho^2}(\dd\eta-v\, \dd\sigma)^2+\dfrac{P}{\rho^2}(\dd\eta+u\,\dd\sigma)^2+\dfrac{\rho^2}{Q}\,\dd q^2+\dfrac{\rho^2}{P}\,\dd p^2\Big],\label{gen_off_sh_Carter}
\end{align}
where ${\rho^2=u+v}$, $u$ and $Q$ are the functions of $q$ only, while $v$ and $P$ are the functions of $p$ only. The case with the null orbits can be obtained by considering the near-horizon limit of the metric (\ref{gen_off_sh_Carter}), see section 2.2.2 in \cite{Ovcharenko2024thesis}.

The corresponding null tetrad for this metric can be found using the relations (\ref{tetrad}) and (\ref{null_tetr_1}), and one can check that for this tetrad ${\kappa=0=\nu}$ and ${\sigma=0=\lambda}$. Also, direct calculation shows that $\Psi_0=0$ in this frame, and thus $\mathbf{k}$ is the PND. The fact that $\mathbf{l}$ is also a PND follows from the symmetry of (\ref{PND_eq}) with respect to the inversion $K\to1/K$ for $\Psi_0=0$ (we remind that in this frame $\Psi_3=\Psi_1$). Thus, to the parameter $K=0$ (meaning that $\mathbf{k}$ is a PND) corresponds a parameter $K\to \infty$ (meaning that $\mathbf{l}$ is also a PND).

Unlike the case considered in \cite{Debever1984}, we have found that most of the conditions on the optical scalars, employed to find the metric (\ref{gen_off_sh_Carter}), hold \emph{automatically} for \emph{any} stationary axisymmetric spacetime of type~D with PNDs orthogonal to the polar direction, \emph{without employing the field equations}. And if one applies the additional conditions that the corresponding PNDs are geodesic and shear-free and that the 1-form $\theta$ is closed, then one automatically obtains the metric (\ref{gen_off_sh_Carter}).

The only remaining thing to prove is that the metric under the requirement of being of algebraic type D becomes (\ref{gen_metr}). The corresponding constraint comes from noticing that for the generic $u$ and $v$, the $\Psi_1$ component is
\begin{align}
    \Psi_1&=\dfrac{\sqrt{PQ}\,\Omega^2}{8\rho^2}\big((\partial_q^2+\partial_p^2)\log\rho^2\big).\label{eq_37}
\end{align}

In Appendix~\ref{app_calcul} we present the detailed calculation showing that the only non-trivial solution to the condition ${\Psi_1=0}$ is
\begin{align}
    u=q^2, \qquad v=p^2.
\end{align}

The same condition was also obtained and integrated in \cite{Debever1984}. However, because of the different notations used in our work and in \cite{Debever1984}, we conduct an analogous calculation employing our notations.

Combining all the results considered in this Section and all the subcases considered in Appendix \ref{app_calcul}, we conclude that in all of them it is possible to bring the metric exactly to the form given in (\ref{gen_metr}), which finishes the proof of Theorem \ref{th_1}.

Several notes related to this theorem should be made. First of all, we would like to emphasize that during this proof we \emph{did not use any field equations}. It means that this theorem holds generally, and the metric ansatz (\ref{gen_metr}) can be used to find exact solutions also in modified theories of gravity.

In addition, we wish to mention that for the metric (\ref{gen_metr}), using the null tetrad (\ref{null_tetr_1}) and the optical scalars for this tetrad calculated in \cite{Ovcharenko2025} (see Eq.~(7) therein), one obtains that the 1-form $\theta$, defined in assumption (v) of Theorem \ref{th_1}, is given by
\begin{align}
    \theta=\dfrac{1}{2}\Big[\Big(\ln \dfrac{\Omega^2}{\rho^2}\Big)_{,q}\dd q+\Big(\ln \dfrac{\Omega^2}{\rho^2}\Big)_{,p}\dd p\Big]+\dfrac{\im}{2}\big[(\ln\rho^2)_{,p}\dd q-(\ln\rho^2)_{,q}\dd p\big].
\end{align}

The real part of this form is the exterior derivative of $\ln (\Omega/\rho)$, and thus the 1-form $\theta$ is given by
\begin{align}
    \theta=\dd\Big(\ln\dfrac{\Omega}{\rho}\Big)+\dfrac{\im}{2}\big[(\ln\rho^2)_{,p}\,\dd q-(\ln\rho^2)_{,q}\,\dd p\big].
\end{align}
Calculating $\dd \theta$, the real part automatically vanishes, while the imaginary part gives
\begin{align}
    \dd \theta=\dfrac{\im}{2}\big[(\ln\rho^2)_{,pp}+(\ln\rho^2)_{,qq}\big]\, \dd p\wedge \dd q.
\end{align}
Taking into account that for the metric (\ref{gen_metr}), \eqref{rho-Q-P-Omega_assump} the function is ${\rho^2=q^2+p^2}$, the whole imaginary part of $\dd\theta$ is also zero, and thus $\theta$ is indeed closed for such a metric.

\section{Proof of Theorem 2: Uniqueness of the non-aligned solution}\label{sec_uniq}

In this section, we will prove that the Ovcharenko-Podolsk\'{y} class of spacetimes, presented in \cite{Ovcharenko2025} as an exact solution of the Einstein-Maxwell equations with a fully non-aligned electromagnetic field, is unique for the ansatz (\ref{gen_metr}). In \cite{Ovcharenko2025} we assumed that the condition
\begin{align}
    \Phi_0=\dfrac{c}{\Omega} \dfrac{\sqrt{P Q}}{q+\im \,p}  \label{Phi_0_relab}
\end{align}
holds for $c=\mathrm{const.}$ To find whether this solution is the most general or not, let us assume that $c$ \emph{is not a constant, but rather a general function of} $q$ \emph{and} $p$ (the fact that $c$ is a function of only $q$ and $p$ follows from the stationarity and axial symmetry),
\begin{align}
    \Phi_0=\dfrac{c(q,p)}{\Omega} \dfrac{\sqrt{P Q}}{q+\im \,p}.
\end{align}

We also use the fact that ${\Phi_0=\Phi_2}$ (this relation is still true as a consequence of stationarity and axial symmetry of the metric (\ref{gen_metr})), and we will prove that necessarily $c=\mathrm{const.}$

To this end, we  employ the Newman-Penrose formalism. We introduce a tetrad (\ref{tetrad}) with ${u=q^2}$ and ${v=p^2}$,
 \begin{align}
    \mathbf{e}_0&=\dfrac{1}{\sqrt{Q}}\dfrac{\Omega}{\rho}(q^2\partial_{\eta}-\partial_{\sigma}),~~~\mathbf{e}_1=\dfrac{\Omega}{\rho}\sqrt{Q}\,\partial_q\\
    \mathbf{e}_3&=\dfrac{1}{\sqrt{P}}\dfrac{\Omega}{\rho}(p^2\partial_{\eta}+\partial_{\sigma}),~~~\mathbf{e}_2=\dfrac{\Omega}{\rho}\sqrt{P}\,\partial_p
\end{align}
and the related null tetrad (\ref{null_tetr_1}).

The Ricci scalars with respect to this null tetrad are given by (see Eqs.~(8)--(12) from \cite{Ovcharenko2025})
\begin{align}
        \Phi_{00}&=\dfrac{Q\Omega}{2\rho^2}\,\Omega_{,qq}=\Phi_{22},\label{Phi_00_02}\\
        \Phi_{02}&=-\dfrac{P\Omega}{2\rho^2}\,\Omega_{,pp}=\bar{\Phi}_{20},\label{Phi_02_20}\\
        \Phi_{01}&=\dfrac{\sqrt{P Q}}{2\rho^4}\Omega\Big(\im\,\rho^2 \,\Omega_{,qp}-(q+\im\,p)(\Omega_{,q}+\im\,\Omega_{,p})\Big)  =\Phi_{21} \nonumber\\
        &=\bar{\Phi}_{10}=\bar{\Phi}_{12}\,,\label{Phi_01}\\
        \Phi_{11}&=\dfrac{1}{8\rho^4}\Bigg[\rho^6\Omega^4\Big(\dfrac{Q_{,q}}{\Omega^2\rho^4}\Big)_{,q}
             +2Q\Big[\big(\Omega^2(\rho^2)_{,q}\big)_{,q}-\rho^2\Omega\Omega_{,qq}\Big]\nonumber\\
    &\hspace{10mm} -\rho^6\Omega^4\Big(\dfrac{P_{,p}}{\Omega^2\rho^4}\Big)_{,p}
             -2P\Big[\big(\Omega^2(\rho^2)_{,p}\big)_{,p}-\rho^2\Omega\Omega_{,pp}\Big]\Bigg],\label{phi_11}
\end{align}
and the Maxwell equations take the form (see Eqs.~(44)--(45) from \cite{Ovcharenko2025})
\begin{align}
    &\sqrt{Q}\,\Bigg[\dfrac{\Omega^2}{\rho^2}\Big(\dfrac{\rho^2}{\Omega^2}\Phi_1\Big)_{,q}-\im\,(\ln \rho^2)_{,p}\, \Phi_1\Bigg]
    +\im\,\Omega^2\Big(\dfrac{\sqrt{P}}{\Omega^2}\Phi_0\Big)_{,p}=0\,,\label{Max_eq_1}\\
    &\sqrt{P}\,\Bigg[\dfrac{\Omega^2}{\rho^2}\Big(\dfrac{\rho^2}{\Omega^2}\Phi_1\Big)_{,p}+\im\,(\ln \rho^2)_{,q}\, \Phi_1\Bigg]
    -\im\,\Omega^2\Big(\dfrac{\sqrt{Q}}{\Omega^2}\Phi_0\Big)_{,q}=0\,.\label{Max_eq_2}
\end{align}

To solve these equations, we introduce a function $ h$ instead of the scalar $\Phi_1$ such that
\begin{align}
    \Phi_1=\dfrac{\Omega^2}{4(q+\im p)^2}\,h(q,p).  \label{Phi_1_relab}
\end{align}
It leads to a great simplification of the Einstein-Maxwell equations ${\Phi_{AB}=2 \Phi_A\bar{\Phi}_B}$. Indeed, substituting (\ref{Phi_0_relab}) and (\ref{Phi_1_relab}) into (\ref{Phi_01}), one can see that it will be satisfied only if
\begin{align}
    h=-\dfrac{1}{\bar{c}} \big[\,\im (q+\im\, p)\Omega_{,qp}+\Omega_q-\im\,\Omega_{,p}\,\big].  \label{h_fun}
\end{align}
This relation directly relates the functions $h$, $c$, and $\Omega$.

As the next step is to solve the Maxwell equations (\ref{Max_eq_1})--(\ref{Max_eq_2}). Substituting (\ref{Phi_1_relab}) and employing the Einstein equations ${\Phi_{00}=2\Phi_0\bar{\Phi}_0}$ and ${\Phi_{02}=2\Phi_0\bar{\Phi}_2}$, using (\ref{Phi_00_02}), (\ref{Phi_01}) to eliminate the functions $P$ and $Q$, the Maxwell equations become
\begin{align}
    h_{,q}+\im (q+\im\,p)^2\Big(\dfrac{\Omega_{,qq}}{\bar{c}\,(q+\im\,p)}\Big)_{,p}=0,\\
    h_{,p}+\im (q+\im\,p)^2\Big(\dfrac{\Omega_{,pp}}{\bar{c}\,(q+\im\,p)}\Big)_{,q}=0.
\end{align}

Substituting the relation (\ref{h_fun}), these equations miraculously simplify and reduce to
\begin{align}
    \Big(\im(q+\im p)\Omega_{,qp}+\Omega_{,q}-\im \Omega_{,p}\Big)\bar{c}_{,q}-\im \Big((q+\im p)\Omega_{,qq}\Big)\bar{c}_{,p}&=0,\label{big_set}\\
    -\im \Big((q+\im p)\Omega_{,pp}\Big)\bar{c}_{,q}+\Big(\im(q+\im p)\Omega_{,qp}+\Omega_{,q}-\im \Omega_{,p}\Big)\bar{c}_{,p}&=0.
\end{align}

This set of equations contains only derivatives of $\bar{c}$. In fact, it is a \emph{linear} system for $\bar{c}_{,q}$ and $\bar{c}_{,p}$ which generally has 2~types of solutions

(i) If
\begin{align}
     D\equiv\Big(\im(q+\im p)\Omega_{,qp}+\Omega_{,q}-\im \Omega_{,p}\Big)^2+(q+\im\,p)^2(\Omega_{,qq})(\Omega_{,pp})\neq 0,
\end{align}
then the solution is trival: ${\bar{c}_{,q}=\bar{c}_{,p}=0\Rightarrow c=\mathrm{const.}}$ In this case, we recover the \emph{Ovcharenko-Podolsk\'{y} class} \cite{Ovcharenko2025}.

(ii) If
\begin{align}
  D=0,  \label{D_cond_0}
\end{align}
the situation is more complicated. In this case, we cannot determine $\bar{c}$ from the Maxwell equations, we can only constrain the $\Omega$ function. We analyze the condition (\ref{D_cond_0}) in detail in Appendix~\ref{sec_solving_field_eqs}, in which we prove that the only non-trivial solution with the non-null electromagnetic field is the \emph{Pleba\'{n}ski-Demia\'{n}ski class} of spacetimes \cite{Plebanski:1976gy} with an \emph{aligned} electromagnetic field.

This finishes the proof.

\section{Conclusions}

We investigated the uniqueness of various exact type D spacetimes in the  Einstein-Maxwell gravity with a non-null electromagnetic field. First, we proved Theorem~\ref{th_1} in which we found that the conformal-to-Carter metric (\ref{gen_metr}), (\ref{rho-Q-P-Omega_assump}) is the most general ansatz satisfying the natural assumptions of Theorem~\ref{th_1}. These conditions include the natural ones, inherited from the realistic Schwarzschild and Kerr black holes (namely stationarity and axial symmetry, being of algebraic type D, and having geodesic and shear-free PNDs), and two additional (that the PNDs are orthogonal to the polar direction~$\partial_\vartheta$, and that the specific 1-form~$\theta$ defined in assumption (v) of Theorem \ref{th_1} is closed). These two additional conditions may seem obscure but they appeared to be vital to obtain the conformal-to-Carter spacetime.

To prove this result we did not employ field equations, and thus the metric ansatz can be used in any modified theory of gravity. We hope that our observation will be used to find new exact solutions.

In the second part of this paper, we applied this result to show in Theorem \ref{th_2} that the Ovcharenko-Podolsk\'{y} class recently found in \cite{Ovcharenko2025} \emph{is the only spacetime} for the conformal-to-Carter ansatz with the fully non-aligned and non-null electromagnetic field in the Einstein-Maxwell theory. In particular, this new result shows that the several assumptions we made in \cite{Ovcharenko2025} to obtain the class of solutions are not required because even without them it is a unique class.

Finally, we wish to mention that even though we have proven that the Ovcharenko-Podolsk\'{y} class \cite{Ovcharenko2025} is quite general, it \emph{does not cover all} type D electro-vacuum spacetimes in the Einstein-Maxwell theory, distinct from the Pleba\'{n}ski-Demia\'{n}ski double-aligned class. In particular, there exist many other exact solutions that are also of type D, but they differ from both Pleba\'{n}ski-Demia\'{n}ski and Ovcharenko-Podolsk\'{y} ones. Namely, there exists the Garc\'{i}a-Pleba\'{n}ski spacetime \cite{Garcia1982} that is of type D, has a fully aligned electromagnetic field, but the PNDs \emph{are not shear-free} (but are still geodesic). Also, there exists the Pleba\'{n}ski-Hacyan solution \cite{Plebanski1979} that is of type D, has a fully aligned electromagnetic field, but the PNDs \emph{are not geodesic} (but are still shear-free). Moreover, there is the Leroy solution \cite{Leroy1978,Leroy1979} of type D, but with only a \emph{partial alignment} of the electromagnetic field with the two PNDs, for which the Kerr spacetime is \emph{not} obtained in the vacuum subcase. Finally, let us mention that there exists a type D spacetime with \emph{null} electromagnetic field found by Debever, Van den Bergh, and Leroy \cite{Debever1989}.  This demonstrates that there are many exact solutions with various interplays between the PNDs of the Weyl type D tensor and the eigendirections of the Faraday tensor. Investigations of the uniqueness of the aforementioned solutions are left for future works.

\subsection*{Supplementary material}
Main expressions and derivations related to this paper are contained in the supplementary
Wolfram Mathematica file \cite{supp_mat}.

\subsection*{Acknowledgements}

This work was supported by the Czech Science Foundation Grant No.~GA\v{C}R 26-22381S, and by the Charles University Grant No.~GAUK 260325.

\appendix

\section{Calculation of $u$ and $v$ functions by solving ${\Psi_1=0}$}\label{app_calcul}

Here we will systematically perform the integration of the condition ${\Psi_1=0}$, with $\Psi_1$ given by (\ref{eq_37}). This means that the condition
\begin{align}
    (\partial_q^2+\partial_p^2)\log\rho^2=0\label{log_eq}
\end{align}
has to be satisfied. The most general solution to this equation is
\begin{align}
    \rho^2=f_1(q+\im p)\,f_2(q-\im p),
\end{align}
where $f_{1,2}$ are general functions. However, we also know from Eq.~(\ref{eq_10}) and Theorem \ref{th_1} that $\rho^2=u(q)+v(p)$. Thus, our task is to find the functions $f_{1}$ and $f_2$, and functions $u$ and $v$, that the functional condition
\begin{align}
    f_1(q+\im p)\,f_2(q-\im p)=u(q)+v(p)\label{funcs_cond}
\end{align}
is satisfied.

To find such functions, let us take the mixed derivative $\dfrac{\partial^2}{\partial q\,\partial p}$ of (\ref{funcs_cond}). The RHS of this equation will be identically zero, while the LHS is not, and it gives
\begin{align}
    \im \big[f_1''(z)f_2(\bar{z})-f_1(z)f_2''(\bar{z})\big]=0.
\end{align}
Here we introduced new coordinates $z=q+\im p,~\bar{z}=q- \im p$, and $'$ denotes a derivative with respect to the corresponding argument. This equation can be rewritten as
\begin{align}
    \dfrac{f_1''(z)}{f_1(z)}=\dfrac{f_2''(\bar{z})}{f_2(\bar{z})}.
\end{align}

As the arguments of the RHS and LHS of this equation are different, they can be equal only if they are constant, namely
\begin{align}
    \dfrac{f_1''}{f_1}=\lambda,~~~\dfrac{f_2''}{f_2}=\lambda.\label{f_12_eqs}
\end{align}
Depending on the value of the separation constant $\lambda$, one gets various results. Let us consider them separately.

\subsection{Case $\lambda=0$}
In this case, (\ref{f_12_eqs}) gives $f_1''=0=f_2''$, which can be easily integrated to
\begin{align}
    f_1=c_0+c_1 (q+\im p),~~f_2=c_2+c_3(q-\im p),
\end{align}
where $c_{i}$ are some constants (generally, complex). Then the function $\rho^2$ is given by
\begin{align}
    \rho^2&=f_1(q+\im p)f_2(q-\im p)=c_0c_2+c_1c_2(q+\im p)+c_0c_3(q-\im p)+c_1c_3 (q^2+p^2) \\
    &=c_0c_2+(c_1c_2+c_0c_3)q+\im p (c_1c_2-c_0c_3)+c_1c_3(q^2+p^2).
\end{align}
However, from the definition the function $\rho^2$ has to be real. This is achieved if
\begin{align}
    \Imm(c_1c_2+c_0c_3)=0,~~\Ree(c_1c_2-c_0c_3)=0,~\Imm(c_0c_2)=0=\Imm(c_1c_3). \label{real_conds}
\end{align}
Introducing a new complex variable $c_1c_2=\alpha$, one can solve the conditions (\ref{real_conds}) that give
\begin{align}
    c_1c_2=\alpha,~c_0c_3=\bar{\alpha},~~c_0c_2=|\alpha|,~~c_1c_3=|\alpha|.
\end{align}
Using these relations, $\rho^2$ becomes
\begin{align}
    \rho^2=|\alpha|+(\alpha+\bar{\alpha})q+\im (\alpha-\bar{\alpha})p+|\alpha|q^2+|\alpha|p^2.
\end{align}

From the other side, $\rho^2$ has to be equal to $\rho^2=u(q)+v(p)$. Comparing these two expressions, one sees that $u$ and $v$ are given by
\begin{align}
    u=u_0+(\alpha+\bar{\alpha})q+|\alpha| q^2,~~v=v_0+\im (\alpha-\bar{\alpha}) p+|\alpha|p^2,
\end{align}
where $u_0$ and $v_0$ are real constants such that $v_0+u_0=|\alpha|$.

Substituting such $u$ and $v$ into the metric (\ref{gen_ax_sym}), one can see that it is not in the form of~(\ref{gen_metr}). However, performing the transformation of  coordinates and relabeling the metric functions as
\begin{align}
    &\eta+u_0\sigma -\dfrac{(\alpha+\bar{\alpha})^2}{4|\alpha|}\sigma=\eta',~~~\sigma=\sigma',\\
    &q+\dfrac{\alpha+\bar{\alpha}}{2|\alpha|}=q',~~~p+\im \dfrac{\alpha-\bar{\alpha}}{2|\alpha|}=p',\\
    &Q=|\alpha|Q',~~~P=|\alpha|P',~~~\Omega'=\Omega,
\end{align}
the metric becomes
\begin{align}
    \dd s^2=\dfrac{1}{\Omega'^2}\Big[-\dfrac{Q'}{\rho'^2}(\dd \eta'-p'^2\dd \sigma')^2
      +\dfrac{P'}{\rho'^2}(\dd \eta'+q'^2\dd \sigma')^2
      +\dfrac{\rho'^2}{Q'}\dd q'^2+\dfrac{\rho'^2}{P'}\dd p'^2\Big],
\end{align}
where ${\rho'^2=q'^2+p'^2}$. This is exactly the metric (\ref{gen_metr}).

\subsection{Case $\lambda$ is real, and $\lambda=-\omega^2<0$}\label{sec_2nd_subcase}

In this case, the general solution to (\ref{f_12_eqs}) is given by
\begin{align}
    f_1=c_0 \cos(\omega z)+c_1 \sin (\omega z),~~~f_2=c_2 \cos(\omega \bar{z})+c_3 \sin(\omega \bar{z}),\label{sol_omega}
\end{align}
and
\begin{align}
    \rho^2=f_1f_2& =\ c_0 c_2 \cos(\omega z) \cos(\omega \bar{z})+c_0c_3\cos(\omega z)\sin(\omega \bar{z})\nonumber\\
    &\,\,\,+c_1c_2\sin(\omega z)\cos(\omega \bar{z})+c_1 c_3 \sin(\omega z)\sin(\omega \bar{z}).
\end{align}

We can simplify these expressions. Using usual trigonometric equations and the definition of $z=q+\im p$, one obtains
\begin{align}
    \cos(\omega z)\cos(\omega \bar{z})&=\dfrac{1}{2}[\cos(2\omega q)+\cosh(2\omega p)],\\
    \cos(\omega z)\sin(\omega \bar{z})&=\dfrac{1}{2}[\sin (2\omega q)-\im \sinh(2\omega p)],\\
    \cos(\omega \bar{z})\sin(\omega z)&=\dfrac{1}{2}[\sin (2\omega q)+\im \sinh(2\omega p)],\\
    \sin(\omega z)\sin(\omega \bar{z})&=\dfrac{1}{2}[-\cos (2\omega q)+ \cosh(2\omega p)],
\end{align}
and $\rho^2$ becomes
\begin{align}
    \rho^2=&\ \big[(c_0c_2-c_1c_3)\cos(2\omega q)+(c_1c_2+c_0c_3)\sin(2\omega q)\nonumber\\
    &\!\! +(c_0c_2+c_1c_3)\cosh(2\omega p)+\im (c_1c_2-c_0c_3)\sinh(2\omega p)\big]/2.\label{rho2_omega}
\end{align}
This expression has to be real, which gives
\begin{align}
    \Imm(c_0c_2)=0,~\Imm(c_1c_3)=0,~\Imm (c_1c_2+c_0c_3)=0,~\Ree(c_1c_2-c_0c_3)=0.
\end{align}
If we introduce two complex constants $\alpha$ and $\beta$ by
\begin{align}
    c_0c_2+\im c_1c_3=\alpha,\quad c_1c_2+c_0c_3=\beta,
\end{align}
then
\begin{align}
    \rho^2=\big[\im (\alpha-\bar{\alpha})\cos(2\omega q)+(\beta+\bar{\beta})\sin(2\omega q)+(\alpha+\bar{\alpha}) \cosh(2\omega p)+\im (\beta-\bar{\beta})\sinh(2\omega p)\big]/2.
\end{align}
By the introduction of suitable new real parameters $a$, $b$, $q_0$, $p_0$, the function $\rho^2$ can be always written in the form
\begin{align}
    \rho^2=a^2 \sin^2[\omega(q+q_0)]+b^2 \sinh^2[\omega(p+p_0)].\label{eq_47}
\end{align}
(Here, we do not write the exact expressions for $a,~b,~q_0$, $p_0$, it is only important that such a transformation can be performed.) However, the function $\rho^2$ given by (\ref{eq_47}) satisfies (\ref{log_eq}) only if $b=a$. Then, employing the fact that $\rho^2=u+v$ we deduce that $u$ and $v$ have to be
\begin{align}
    u&=u_0+a^2 \sin^2[\omega(q+q_0)],\\
    v&=-u_0+a^2 \sinh^2[\omega(p+p_0)].
\end{align}

It may again seem that these $u$ and $v$ do not give rise to the metric (\ref{gen_metr}). However, if one performs the transformation
\begin{align}
    &q=\dfrac{1}{\omega}\arcsin\Big(\dfrac{q'}{\sqrt{a^2+q'^2}}\Big)-q_0,~~~p=\dfrac{1}{\omega}\mathrm{arcsinh}\Big(\dfrac{p'}{\sqrt{a^2-p'^2}}\Big)-p_0,\\
    &\eta+u_0\sigma=\eta',~~~(a^2+u_0)\sigma+\eta=\sigma',\\
    &Q'=\dfrac{a}{\omega}Q(a^2+q'^2)^2,~~~P'=\dfrac{a}{\omega}P(a^2-p'^2)^2,\\
    &\Omega'^2=\Omega^2 \dfrac{\omega}{a}(a^2-p'^2)(a^2+q'^2),
\end{align}
then the metric (\ref{gen_ax_sym}) becomes
\begin{align}
    \dd s^2=\dfrac{1}{\Omega'^2}\Big[-\dfrac{Q'}{\rho'^2}(\dd \eta'-p'^2\dd \sigma')^2
    +\dfrac{P'}{\rho'^2}(\dd \eta'+q'^2\dd \sigma')^2
    +\dfrac{\rho'^2}{Q'}\dd q'^2+\dfrac{\rho'^2}{P'}\dd p'^2\Big],
\end{align}
where $\rho'^2=q'^2+p'^2$, exactly the metric (\ref{gen_metr}).

\subsection{Case $\lambda$ is real, and $\lambda=\omega^2>0$}

The proof in this case is analogous, one only has to change all trigonometric functions to hyperbolic, and vice versa. As this transition is clear, we do not repeat a systematic derivation for this case.

\subsection{Case $\lambda$ is complex}

Finally, let us consider the most general case, namely when $\lambda$ is complex. In this case, we have the freedom to introduce a new quantity $\omega$ such that $\lambda=-\omega^2$. However, unlike the case considered in the Sec.~\ref{sec_2nd_subcase}, now $\omega$ is a complex number, namely
\begin{align}
    \omega=\omega_R+\im\, \omega_I.\label{compl_omega}
\end{align}

For the complex $\omega$, relations (\ref{sol_omega}) still hold. This means that generally the relation (\ref{rho2_omega}) also holds. Let us analyze the $q$-dependent terms in this relation and require them to be real. First of all, we notice that, using (\ref{compl_omega}), the $q$-dependent terms become
\begin{align}
    \cos(2\omega q)&=\cos(2\omega_R q)\cosh(2\omega_I q)-\im \sin(2\omega_R q)\sinh(2\omega_I q),\\
    \sin(2\omega q)&=\sin(2\omega_R q)\cosh(2\omega_I q)+\im \cos(2\omega_R q)\sinh(2\omega_I q).
\end{align}
Thus, the $q$-dependent part in (\ref{rho2_omega}) becomes
\begin{align}
    &\quad\, (c_0c_2-c_1c_3)\cos(2\omega_R q)\cosh(2\omega_I q)-\im (c_0c_2-c_1c_3)\sin(2\omega_R q)\sinh(2\omega_I q)\nonumber\\
    &+(c_1c_2+c_0c_3)\sin(2\omega_R q)\cosh(2\omega_I q)+\im (c_1c_2+c_0c_3)\cos(2\omega_R q)\sinh(2\omega_I q).
\end{align}

All four terms represent different functional dependence on $q$, thus these terms can be made real \emph{only} if all the coefficients are real. However, looking at them, one deduces that it can be obtained only if the corresponding coefficients are zero. This means that all the terms depending on~$q$ have to be absent in $\rho^2$. In the same manner, we show that the terms depending on~$p$ are also zero. This gives the trivial solution $\rho^2=0$, that must be omitted.

\section{Solving the field equations}\label{sec_solving_field_eqs}

First of all, let us notice that the condition
\begin{align}
     D\equiv\Big(\im(q+\im p)\Omega_{,qp}+\Omega_{,q}-\im \Omega_{,p}\Big)^2+(q+\im\,p)^2(\Omega_{,qq})(\Omega_{,pp})= 0\label{D_cond}
\end{align}
is complex, while $\Omega$ is assumed to be a real function. This means that the two conditions for the real and imaginary parts have to be satisfied simultaneously. To proceed further, let us thus consider the specific combination
\begin{align}
    q p \,\Ree(D)-\dfrac{(q^2-p^2)}{2}\Imm(D)=(q\, \Omega_{,q}-p\,\Omega_{,p})\Big((q^2+p^2)\,\Omega_{,qp}-q\,\Omega_{,p}-p\,\Omega_{,q}\Big). \label{fin_cond}
\end{align}
This expression is factorized into two brackets. Each of them can be zero. Let us consider both these cases separately.

\subsection{1st case}
In the first case, let us assume that the first bracket in (\ref{fin_cond}) is zero, namely that
\begin{align}
    q\, \Omega_{,q}-p\,\Omega_{,p}=0.
\end{align}
This equation is quite simple, and its general solution is such that $\Omega$ is the function of the product $qp$ solely, namely ${\Omega=\Omega(qp)}$. By the direct substitution, one can check that for such a form of $\Omega$ the condition $D=0$ is automatically satisfied.

Now, let us solve the remaining field equations with this assumption. We start with the $\Phi_{00}$ and $\Phi_{02}$ equations. As for stationary axisymmetric spacetimes $\Phi_0=\Phi_2$, it directly follows from (\ref{Phi_00_02}) that $\Phi_{00}-\Phi_{02}=0$. Substituting ${\Omega=\Omega(qp)}$, one obtains that this equation takes the form
\begin{align}
    \Phi_{00}-\Phi_{02}=\Omega\Omega''\,\dfrac{Pq^2+p^2 Q}{2(q^2+p^2)}.
\end{align}
This equation has two possible solutions: either ${\Omega''=0}$, or ${Pq^2+Qp^2=0}$.

In the case when $\Omega''=0$, it follows that ${\Phi_{00}=0=\Phi_{02}}$, meaning that the non-aligned part of the electromagnetic field has to vanish. In addition, integrating ${\Omega''=0}$ one obtains ${\Omega=a+b\,qp}$, where $a$ and $b$ are constants. As can be seen, this case corresponds to the \emph{Pleba\'{n}ski-Demia\'{n}ski spacetime} with an aligned electromagnetic field.

Now let us focus on the second possibility, $Pq^2+Qp^2=0$. This condition can be satisfied only if
\begin{align}
    Q=\beta q^2,~~~P=-\beta p^2,~~~\beta=\mathrm{const.}
\end{align}
As can be checked by a substitution of $\Omega=\Omega(qp)$ into (\ref{h_fun}), and using (\ref{Phi_1_relab}) and equations (\ref{Phi_00_02}), one finds that
\begin{align}
    \Phi_1=\im \,\beta \dfrac{c\,q\,p}{\Omega(q+\im\,p)}.
\end{align}
(Here and further, we assume that $\beta$ is positive. The proof for a negative $\beta$ is analogous.)
From the other side, the $\Phi_{0}$ component defined in (\ref{Phi_0_relab}) is equal to
\begin{align}
    \Phi_0=\im \,\beta \dfrac{c\,q\,p}{\Omega(q+\im\,p)}.
\end{align}
It means that for this special subcase ${\Phi_1=\Phi_0}$, meaning that the \emph{electromagnetic field is null}. However, we can even prove that this case is \emph{incompatible} with the rest of the field equations. To see this, we note that the condition ${2\Phi_1\bar{\Phi}_1=\Phi_{11}=\Phi_{00}=2\Phi_{0}\bar{\Phi}_0}$ has to hold because ${\Phi_1=\Phi_0}$. However, a direct substitution of $\Omega=\Omega(qp),~P=-\beta p^2,~Q=\beta q^2$ into (\ref{Phi_00_02}) and (\ref{phi_11}) gives us
\begin{align}
    \Phi_{00}= \beta \Omega\, \dfrac{q^2p^2 \Omega''}{2(q^2+p^2)},~~~
    \Phi_{11}=-\beta \Omega\  \dfrac{q^2p^2 \Omega''}{2(q^2+p^2)}.
\end{align}
It is now directly seen that the equation ${\Phi_{00}=\Phi_{11}}$ can \emph{only} be satisfied if ${\beta=0}$ (unless ${\Omega''=0}$ which is the Pleba\'{n}ski-Demia\'{n}ski spacetime), giving the trivial ${P=0=Q}$ solution.

To conclude, in the first case we have considered here, there is only one non-trivial solution, and it is the Pleba\'{n}ski-Demia\'{n}ski one.

\subsection{2nd case}

Now let us consider the second case, namely that the second bracket in (\ref{fin_cond}) is zero,
\begin{align}
    (q^2+p^2)\,\Omega_{,qp}=q\,\Omega_{,p}+p\,\Omega_{,q}.\label{eq_1}
\end{align}
This equation is quite complicated to solve. However, we have to require both real and imaginary parts of (\ref{D_cond}) to be zero. Under the assumption (\ref{eq_1}), the real part of (\ref{fin_cond}) simplifies and becomes
\begin{align}
    -\dfrac{(q^2+p^2)^2}{q^2-p^2}\,\Ree{(D)}=(q\Omega_{,q}-p\Omega_{,p})^2+(q^2+p^2)^2(\Omega_{,qq})(\Omega_{,pp})=0.\label{eq_2}
\end{align}
It means that we have to \emph{simultaneously} solve both equations (\ref{eq_1}) and (\ref{eq_2}). Now let us consider the expression for $h$ given by (\ref{h_fun}). This function can be rewritten as
\begin{align}
    h&=-\dfrac{1}{\bar{c}}\big[\im(q+\im\, p)\Omega_{,qp}+\Omega_{,q}-\im \Omega_{,p}\big]=\\
    &=-\dfrac{1}{\bar{c}\,(q-\im p)}\big[\im(q^2+p^2)\Omega_{,qp}+q\Omega_{,q}-p \Omega_{,p}
    -\im (q\Omega_{,p}+p\Omega_{,q})\big].
\end{align}

Notice that the imaginary part of the expression vanishes because of (\ref{eq_1}). The real part can be, using (\ref{eq_2}), written as
\begin{align}
    h=-\dfrac{\im}{\bar{c}\,(q-\im p)}(q^2+p^2)\sqrt{\Omega_{,qq}\Omega_{,pp}}\,,
\end{align}
and thus the whole $\Phi_1$, using (\ref{Phi_1_relab}), can be written as
\begin{align}
    \Phi_1=-\im \dfrac{\Omega^2}{4\bar{c}\,(q+\im p)}\sqrt{\Omega_{,qq}\Omega_{,pp}}\,.\label{Phi_1_simpl}
\end{align}

Let us simplify this expression. Using Eqs.~(\ref{Phi_00_02}) and (\ref{Phi_0_relab}), from the Einstein equations ${\Phi_{00}=2\Phi_0\bar{\Phi}_0}$ and ${\Phi_{02}=2\Phi_0\bar{\Phi}_2}$ one finds that
\begin{align}
    \Omega_{,qq}=4 c \bar{c}\, \dfrac{P}{\Omega^3},~~~
    \Omega_{,pp}=-4 c\, \bar{c} \dfrac{Q}{\Omega^3}.
\end{align}
Substituting these expressions into (\ref{Phi_1_simpl}), one obtains
\begin{align}
    \Phi_1= \dfrac{c\,\sqrt{PQ}}{\Omega(q+\im p)}.
\end{align}
Comparing it with (\ref{Phi_0_relab}), we see that for this case
\begin{align}
    \Phi_1=\Phi_0=\Phi_2.
\end{align}
Calculating the electromagnetic invariant, we get
\begin{align}
    \dfrac{1}{16}F_{\mu\nu}^{*}F^{*\mu\nu}=\Phi_0\Phi_2-\Phi_1^2=0,
\end{align}
meaning that in this case the \emph{electromagnetic field is null}.

The fact that the electromagnetic field is null helps us to prove that this setup is incompatible with the field equations. First of all, as ${\Phi_0\neq 0}$ and ${\Phi_2\neq 0}$, we deduce that the eigendirections of the Faraday tensor are \emph{not aligned} with any of the PNDs of the Weyl tensor. However, this statement is in contradiction with already proven theorems that \emph{null electromagnetic fields are necessarily aligned with the PNDs of the Weyl tensor}, see \cite{Debever1989, Kuchynka2017}. It means that the solution to (\ref{eq_1})--(\ref{eq_2}) is \emph{incompatible with the remainong field equations}. This also implies that the only non-trivial solution to (\ref{big_set}) is $c=\mathrm{const.}$, giving the Ovcharenko-Podolsk\'{y} class.

\end{document}